\documentclass[fleqn,usenatbib]{mnras}

\usepackage{newtxtext,newtxmath}

\usepackage[T1]{fontenc}

\DeclareRobustCommand{\VAN}[3]{#2}
\let\VANthebibliography\thebibliography
\def\thebibliography{\DeclareRobustCommand{\VAN}[3]{##3}\VANthebibliography}

\usepackage{graphicx}	
\usepackage{amsmath}	
\usepackage{amssymb}	
\usepackage{booktabs}   
\usepackage{multirow}   
\usepackage{comment}    
\usepackage[utf8]{inputenc} 

\title[Nature of resolved main sequence]{The ALMaQUEST Survey IX: The nature of the resolved star forming main sequence}
\author[W. M. Baker et al.]
{William M. Baker$^{1,2}$\thanks{E-mail: wb308@cam.ac.uk},
Roberto Maiolino$^{1,2}$,
Asa F. L. Bluck$^{1,2}$,
Lihwai Lin$^{3}$,
Sara L. Ellison$^{4}$,
\newauthor{}
Francesco Belfiore$^{5}$,
Hsi-An Pan$^{6}$,
Mallory Thorp$^{4}$
\\
$^{1}$Kavli Institute for Cosmology, University of Cambridge, Madingley Road, Cambridge, CB3 OHA, UK\\
$^{2}$Cavendish Laboratory - Astrophysics Group, University of Cambridge, 19 JJ Thompson Avenue, Cambridge, CB3 OHE, UK\\
$^{3}$Institute of Astronomy \& Astrophysics, Academia Sinica, Taipei 10617, Taiwan\\
$^{4}$Department of Physics \& Astronomy, University of Victoria, Finnerty Road, Victoria, British Columbia, V8P 1A1, Canada \\
$^{5}$INAF— Osservatorio Astrofisico di Arcetri, Largo E. Fermi 5, I-50125, Florence, Italy\\
$^{6}$Department of Physics, Tamkang University, Tamsui Dist., New Taipei City 251301, Taiwan\\
}

\date{Accepted XXX. Received YYY; in original form ZZZ}

\pubyear{2021}

\begin{document}
\label{firstpage}
\pagerange{\pageref{firstpage}--\pageref{lastpage}}
\maketitle

\begin{abstract}
We investigate the nature of the scaling relations between the surface density of star formation rate ($\Sigma _{\rm SFR}$), stellar mass  ($\Sigma _*$), and molecular gas mass ($\Sigma _{\rm H_2}$), aiming at distinguishing between the relations that are primary, i.e. more fundamental, and those which are instead an indirect by-product of the other relations. We use the ALMaQUEST survey and analyse the data by using both partial correlations and Random Forest regression techniques. We unambiguously find that the strongest intrinsic correlation is between $\Sigma _{\rm SFR}$ and $\Sigma_{\rm H_2}$ (i.e. the resolved Schmidt-Kennicutt relation), followed by the correlation between $\Sigma _{\rm H_2}$ and $\Sigma _*$ (resolved Molecular Gas Main Sequence, rMGMS). Once these two correlations are taken into account, we find that there is no evidence for any intrinsic correlation between $\Sigma _{\rm SFR}$ and $\Sigma _*$, implying that SFR is entirely driven by the amount of molecular gas, while its dependence on stellar mass (i.e. the resolved Star Forming Main Sequence, rSFMS) simply emerges as a consequence of the relationship between molecular gas and stellar mass.
\end{abstract}

\begin{keywords}
Galaxies: formation, evolution, fundamental parameters; star formation
\end{keywords}



\section{Introduction}
Star formation rate is observed to correlate strongly with both stellar mass  \citep{2004MNRASBrinchmann, Whitaker2012ApJ...754L..29W, Renzini+PengMS2015ApJ...801L..29R} and molecular gas mass \citep{Schmidt1959,Kennicutt1998}, both for the integrated quantities and also on spatially resolved scales.
It is important to understand which of these correlations are the strongest and most important, and which of them are secondary relations possibly resulting from the others.

The resolved Schmidt-Kennicutt (rSK) \citep{Schmidt1959,Kennicutt1998} relation   connects molecular gas mass surface density, $\Sigma_{\rm H_2}$, and the star formation rate surface density, $\Sigma_{\rm SFR}$ \citep{2008BigielAJ....136.2846B, 2013LeroyAJ....146...19L, 2017UtomoApJ...849...26U}.
In particular, it quantifies how an increase in resolved molecular gas mass results in an increase in star formation rate. The Schmidt-Kennicutt relation is well understood, to a first order, in terms of molecular gas providing the fuel for star formation \citep{Kennicutt1998}.

The Star Forming Main Sequence (SFMS) is an empirically derived relation which connects stellar mass and star formation rate, both globally \citep{2004MNRASBrinchmann}, between $\rm M_*$ and SFR, 
and locally  \citep{2013SanchezA&A...554A..58S,2016Cano-diasApJ...821L..26C,2017Hsieh, Enia2020MNRAS.493.4107E}, between $\Sigma_*$ and $\Sigma_{\rm SFR}$. This correlation is typically found to be close to linear, or slightly sub-linear.

A relation which has been highlighted more recently is the resolved Molecular Gas Main Sequence (rMGMS), as found in \citet{Lin2019} and \cite{Ellison2021AlmaQuest5} (for star forming regions), and then extended in \citet{Ellison2021AlmaQuest6} and \citet{Lin2021submitted} to passive regions and green valley galaxies. Evidence for the existence of the rMGMS was also found in \cite{2020MorselliMNRAS.496.4606M, Pessa2021A&A...650A.134P,2020BarreraMNRAS.492.2651B}.
The rMGMS connects the molecular gas mass surface density and stellar mass surface density.

Clearly, the three relations discussed above, between $\Sigma _{\rm SFR}$, $\Sigma _*$ and $\Sigma _{\rm H_2}$, are interconnected with each other. This leads to the question: are all three of these relations equally important and  associated with physical processes, or are some of them a by-product of the others?
\citet{Lin2019} investigated this question on spatially resolved scales by analysing star forming regions in 14 main sequence galaxies selected from the ALMA-MaNGA QUEnching and STar formation (ALMaQUEST) survey \citep{Lin2020}. They measured the Pearson correlation coefficients between each two of the three quantities and their scatter. They found that, in terms of correlation strength and scatter, the rSK relation was the strongest.  After the rSK relation, the next most important scaling relation was found by them to be the rMGMS, whilst the least important relationship was found to be the rSFMS.

They suggest that the form of the relations among the resolved quantities ($\Sigma_{\rm SFR}$, $\Sigma_*$ and $\Sigma_{\rm H_2}$) are all essentially the same, and that the three parameters ($\Sigma_{\rm SFR}$, $\Sigma_*$ and $\Sigma_{\rm H_2}$) can be characterised by a line (or cylinder) in 3D log-log space, with the individual relations (rSK, rSFMS, and rMGMS) being simply 2D projections of this 3D relation. This result is also confirmed by the findings of \citep{2021SanchezMNRAS.503.1615S}.

\citet{Lin2019} suggested a scenario to explain the rMGMS in which molecular gas is accreted (or retained) preferentially in regions of deeper gravitational potential. These regions may be dominated by the stellar mass, or both the stellar mass and molecular gas could be tracing deeper gravitational potential regions associated with dark matter.
\citet{Lin2019} also proposed that the rSFMS is a by-product of the rSK and rMGMS relations, as suggested by the fact that the rSFMS has the largest scatter and smallest Pearson correlation coefficient. 
\citet{Ellison2021AlmaQuest5} subsequently confirmed the small scatters and stronger correlations of the rSK and rMGMS on a larger sample, a result also supported by independent datasets \citep{2020MorselliMNRAS.496.4606M}. The resolved scaling relations have also been observed at higher spatial resolutions, such as in \citet{Pessa2021A&A...650A.134P}, where they explored scales as small as $100\text{pc}$. They found that the scaling relations were recovered at these scales.

In \citet{Ellison2021AlmaQuest6} the investigation of the rMGMS is extended to the passive, retired spaxels and comparison is made between the star forming and non star forming spaxels. They found that there was a resolved Molecular Gas Main Sequence for the retired spaxels which was offset from that of the star forming spaxels. Specifically, for a given $\Sigma_*$ the corresponding $\Sigma_{\rm H_2}$ is lower for the retired spaxels than for the star forming spaxels. They explained this via a reduction of gas content leading to lower star formation rates, through the rSK relation, for the retired spaxels.

\citet{2020EllisonSFEMNRAS.493L..39E} explored the scaling relations using correlation coefficients and an artificial neural network, finding that, in agreement with \citet{Lin2019}, the most important parameter for determining $\Sigma_{\rm SFR}$ was $\Sigma_{\rm H_2}$. 
They also found that the scatter in (i.e., the variation above and below) the SFMS was driven primarily by changes in the star formation efficiency ($\Delta$SFE), with the gas fraction having a secondary effect. This suggested that it is the star formation rate relative to the amount of gas that is important rather than simply the amount of gas. 

In this paper we further investigate the nature of, and the connection between, these scaling relations, and in particular the claims of \citet{Lin2019} and \cite{2020EllisonSFEMNRAS.493L..39E}, by using two statistically strong, independent-methods utilising the homogeneous, spatially resolved information of the full ALMaQUEST sample of 46 local galaxies. We select star forming regions regardless of the parent galaxy classification. We determine the strength of the correlations using Partial Correlation Coefficients. 
These enable us to avoid the difficulties involved in measuring intrinsic correlations between three interconnected quantities.
We also use the machine learning based Random Forest method, which enables us to simultaneously investigate the importance of different galactic properties in determining a target quantity, specifically $\Sigma_{\rm SFR}$, without any assumption on the linearity or monotonicity of the relations. This also enables us to assess the relative performance of each parameter in determining $\Sigma_{\rm SFR}$ considering a wide variety of alternative parameters simultaneously.


We are aware that our new statistical analysis confirms the conclusions in \citet{Lin2019} and \cite{2020EllisonSFEMNRAS.493L..39E}, but we place these on a more robust statistical footing. More explicitly, we unambiguously rule out the possibility for measurement errors or non-linearity of the relations impacting the key insight, i.e. that the rSFMS is an emergent relation arising purely out of logical necessity from two more fundamental scaling laws - the rSK and rMGMS relations.

Throughout the paper we assume $H_0=70$km $\text{s}^{-1}$ $\text{Mpc}^{-1}$, $\Omega_m=0.3$ and $\Omega_{\Lambda}=0.7$.

\section{Data \& Methods}

\subsection{MaNGA}

We use the publicly released (DR15), resolved, optical spectroscopic data obtained by the MaNGA (Mapping Nearby Galaxies at Apache Point Observatory) survey \citep{Bundy2015} \footnote{https://www.sdss.org/dr15/manga/manga-data/} analysed through
the PIPE3D pipeline \citep{Sanchez2016}. 
The MaNGA survey was part of the fourth-generation of the Sloan Digital Sky Survey (SDSS) \citep{Blanton2017AJ....154...28B} and used multiple fibre bundles to obtain spatially resolved spectra of approximately 10,000 local ($z\sim0.03$) galaxies with a spectral resolution R$\sim$2000 across the  spectral range $\lambda$3600--10300\AA. The sample was designed to be representative of the local galaxy population, by including both star forming main sequence and quiescent galaxies and selected to have a flat stellar mass distribution between $\rm 10^9M_{\odot}$ and $\rm 10^{11}M_{\odot}$. Two samples were identified, one covering galaxies out to 1.5 effective radii (1.5$R_e$) and a second one covering galaxies out to 2.5 effective radii (2.5$R_e$). No selection was done on galaxy morphology \citep[further details about the survey are given in ][]{Bundy2015,Wake2017AJ....154...86W,Yan2016AJ....152..197Y,Law2016AJ....152...83L}.

From the data products we primarily use the spectral emission line fluxes, stellar mass per spaxel (derived from spectral fitting with SSP), and H$\alpha$ equivalent widths. 
From the PIPE3D Value Added Catalogue we use the total stellar mass per galaxy.
The NASA-Sloan Atlas provides us with the galaxy's redshifts, inclination, and the galactocentric distance of each spaxel (as measured from the centre of their respective galaxy).

\subsection{ALMaQUEST}
We use data from the ALMaQUEST\footnote{For more information see https://arc.phys.uvic.ca/~almaquest/} survey \citep{Lin2020}, which is a survey performed with the Atacama Large Millimetre Array (ALMA), designed to obtain CO(1-0) 115GHz measurements from 46 local galaxies that are included in the MaNGA survey \citep{Bundy2015}. The galaxies were chosen to span a broad range, in terms of specific star formation rate (sSFR), across the main sequence, including green valley and starburst galaxies. The angular resolution of the ALMA measurements is around 2.5'', matching that of the MaNGA observations, and the sensitivity of the CO line flux ranges from about 0.02 to 0.1 Jy km s$^{-1}$ beam$^{-1}$. A detailed description of the ALMaQUEST survey and of its observational parameters is given in \citet{Lin2020}.

\subsection{Derivation of the physical quantities}
We choose spaxels with S/N>2 in all emission lines, H$\alpha$, H$\beta$, [OIII]$\lambda$5007, [NII]$\lambda$6584, [SII]$\lambda$6717,31 and CO(1-0). This enables us to maximise the number of spaxels available for analysis. We test the stability of the results by applying higher signal to noise cuts, and find no significant differences. 
We also only select spaxels with a H$\alpha$ equivalent width greater than 6 Angstroms (EW(H$\alpha$)>6\AA) to minimise possible contamination from diffuse ionised gas. We note that this selection tends to avoid spaxels with low sSFR, relative to the resolved Star Forming Main Sequence, however this is acceptable in this analysis as our focus is to investigate the scaling relations on the Main Sequences, and in particular the origin of the rSFMS.

As we are investigating the scaling relations for star forming regions we need to separate them from the other regions of the galaxies. In addition to the EW constraint discussed above, 
star forming regions are selected using the [NII]--BPT diagram \citep{Baldwin}, with the \citet{2003Kauffmann} dividing line for star forming regions. The specific BPT requirement is also needed to determine the metallicity (as metallicity diagnostics are well calibrated only for star forming regions), which will be required when using the metallicity-dependent CO conversion factor. Note that these criteria select HII regions, i.e. star forming spaxels, regardless of the classification of the parent galaxy  they belonged to, e.g. we are also selecting SF spaxels in Green Valley galaxies (as well as in starburst galaxies). 

The flux of nebular emission lines are corrected for dust extinction by 
using the  \citet{Cardelli1989} extinction curve, with $R_V=3.1$, and the Balmer decrement $F_{H\alpha}/F_{H\beta}$ to estimate the dust reddening (assuming an intrinsic ratio of 2.86, corresponding to case B recombination, and a temperature of $10^4$ K). The extinction-corrected H$\alpha$ flux is then used to determine the star formation rate using the relationship from \citet{1998KennicuttSFR} (assuming a Salpeter IMF)
\begin{equation}
    \mathrm{SFR}[\text{M}_\odot \text{yr}^{-1}]=7.9\times10^{-42}L_{\rm H\alpha}[\text{erg}\, \text{s}^{-1}].
\end{equation}

Metallicities of the spaxels are calculated using the empirical strong line calibrations given in \citet{2017MNRASCurti} and specifically using the following diagnostic ratios,
$\rm R3=log([OIII]5007/H\beta$), $\rm N2 = log([NII]6584/H\alpha)$, and $\rm S2=log([SII]6717,31/H\alpha)$.

To determine the molecular gas mass we convert the CO flux into a CO luminosity following the process detailed in \citet{Solomon1997}. We are then able to calculate the molecular gas mass by applying a constant Milky-Way-like CO conversion factor as in \citet{Lin2020}, defined as \citep{Bolatto2013ARA&A..51..207B}
\begin{equation}
    \alpha_{\rm CO}=4.35\, \text{M}_\odot (\text{K km/s pc}^2)^{-1}.
    \label{eq:con}
\end{equation}
We also test a metallicity-dependent conversion factor, as in \citet{Lin2020}, given by \citep{Sun2020ApJ...892..148S}
\begin{equation}
    \alpha_{\rm CO}=4.35\,\bigg(\frac{Z}{Z_\odot}\bigg)^{-1.6} \text{M}_\odot (\text{K km/s pc}^2)^{-1},
\end{equation}
where $Z/Z_\odot$ is the gas metallicity relative the solar value.
We find that the metallicity-dependent conversion factor does not alter our results significantly and we note that the general results remains the same for either conversion factor.

A galaxy inclination cut is applied which removes any star forming regions belonging to highly inclined galaxies; this is to avoid significant uncertainties associated with strong projection effects when estimating the surface densities and also complex dust extinction effects. We remove spaxels belonging to galaxies with axial ratios $\frac{b}{a}<0.35$, replicating the criterion used in \citet{Ellison2021AlmaQuest5}. 
 We tested a stricter inclination cut of removing spaxels belonging to galaxies with axial ratios $\frac{b}{a}$<0.5 and found that it had no significant effect on our results, hence reassuring us that inclination effects (at least for galaxies with b/a>0.35) are not significantly affecting our results.

We correct all surface density quantities for inclination, such that
\begin{equation}
    \text{log}(\Sigma)_{\text{corrected}}=\text{log}(\Sigma)_{\text{observed}}+
    \text{log}\bigg(\frac{b}{a}\bigg),
\end{equation}
which does not significantly effect any of our results.

We then apply a stellar mass surface density cut of log($\Sigma_* [\rm M_{\odot}/kpc^2]$)>7 to the remaining spaxels, in order to remove the few spaxels whose stellar continuum is so faint that their stellar mass surface density was estimated with significant uncertainty.
In total this gives us 17,072 star forming spaxels belonging to 36 galaxies. In our later analysis, we remove very few outliers which have measurements that are deviating by more than 3$\sigma$ from the mean ($\mu$) for each quantity (assuming a Gaussian distribution) and which are likely associated with observational artifacts. We also test our results by not including the outlier cut, finding that whether the outlier cut is included or not, does not impact our findings.

\section{Results}

\subsection{Partial Correlation Coefficient Results}

\begin{figure}
    \centering
    \includegraphics[width=\columnwidth]{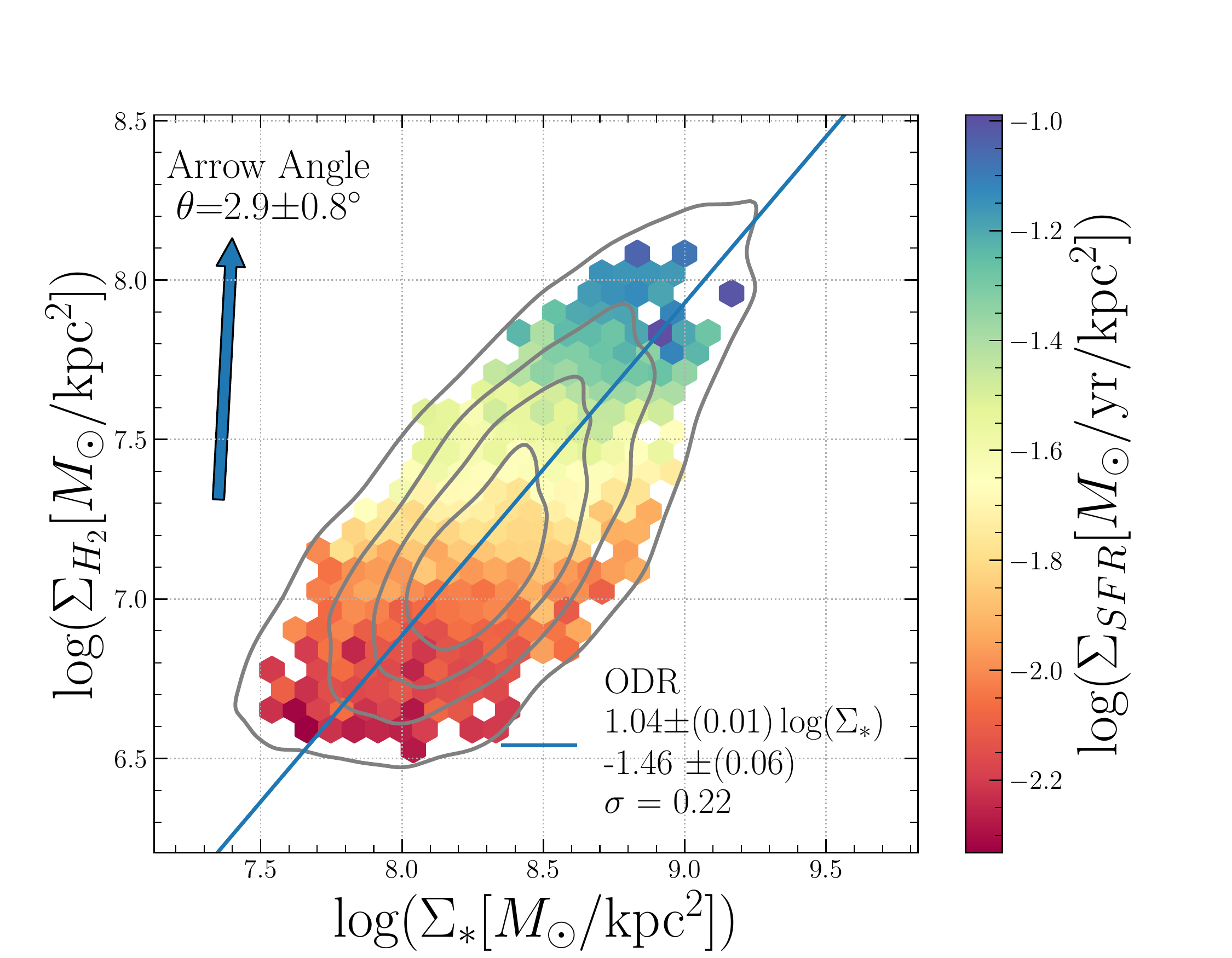}
    \caption{The molecular gas surface density vs the stellar mass surface density, colour-coded by the star formation rate surface density. The partial correlation coefficient arrow points in the direction of the steepest gradient in the $z$ colour-bar quantity. The arrow angle is given by $\theta$ (defined in eq. \ref{theta}). The blue line is the best fit of the $\Sigma_{\rm H_2}-\Sigma_{*}$ relation calculated using ODR. The contours give the density of spaxels, with the outer contour enclosing 90\% of the spaxels sample.}
    \label{fig:1}
\end{figure}

Partial Correlation Coefficients allow us to disentangle the relationships between three quantities that all appear correlated with each other, in a way that other techniques, such as Pearson's r, do not. 
More specifically, Partial Correlation Coefficients allow the investigation of correlations between two quantities whilst keeping constant (controlled) the third quantity, hence enabling us to inspect the true correlations between two quantities that are not driven by their correlation with the third quantity. Quantitatively,
the partial correlation coefficient between two quantities A and B, whilst controlling for quantity C, is given by
\begin{equation}
    \rho_{AB|C}=\frac{\rho_{AB}-\rho_{AC}\rho_{BC}}{\sqrt{1-\rho_{AC}^2}\sqrt{1-\rho_{BC}^2}}
\end{equation}
as in \cite{PCC10.2307/2683864}, where $\rho_{ij}$ is the Spearmann rank correlation coefficient which is calculated between the quantities $i$ and $j$. We note that a monotonic relation between quantities is a requirement for the Partial Correlation Coefficients to be meaningful, and that this is approximately true in our data.


Figure \ref{fig:1} shows the plot for $\Sigma_*$ vs $\Sigma_{\rm H_2}$ with the mean $\Sigma_{\rm SFR}$ colour-coded on the $z$ axis.  The contours in Figure \ref{fig:1} show the density of spaxels on the diagram. This is a form of 3-dimensional plot which, as we will discuss more in detail below,  illustrates the three scaling relations discussed in Section 1 simultaneously (i.e. rSK, rSFMS and rMGMS). 
 This representation already qualitatively and visually illustrates that, 
at a fixed $\Sigma_*$, $\Sigma_{\rm SFR}$ varies significantly with $\Sigma_{H_2}$. However, at a fixed $\Sigma_{H_2}$, $\Sigma_{\rm SFR}$ varies very little (or not at all) with $\Sigma_*$ (i.e. the colour gradient is nearly vertical). Therefore, visually, this illustrates that the dependence between $\Sigma_{\rm SFR}$ and $\Sigma_{H_2}$ is almost independent of $\Sigma_*$; yet, the dependence of $\Sigma_{\rm SFR}$ on $\Sigma_*$ is highly dependent on (and actually nearly totally driven by) $\Sigma_{H_2}$. 

For a more quantitative analysis in this plot, the Partial Correlation Coefficients can be used to determine an arrow that points in the direction of the largest gradient in the $z$ colour-coded quantity.
The equation for determining the angle in a diagram, with $A$, $B$, being the quantities on the $x$ and $y$ axis, respectively, and $C$ being the quantity on the $z$ axis, is given by 
\citep[adapting equation from][]{2020Bluck, Piotrowska2019}
\begin{equation}
    \text{tan}(\theta)=\frac{\rho_{AC|B}}{\rho_{BC|A}}
    \label{theta}
\end{equation}
where the angle $\theta$ is measured from the positive $y$ axis (in a clockwise orientation). The errors on the angle $\theta$ are computed via bootstrap random sampling, where we take 100 random samples of the data with replacement and compute their standard deviation (with the sample size equal to the dataset).

Note that the colour-coding technique and partial correlation coefficient arrow are useful to determine the dependence of the third (colour-coded) quantity on the other two, $x$ and $y$ quantities. However, these techniques should {\it not} be used to determine the dependence of either the $x$ or $y$ quantities on the third $z$ (colour-coded) quantity; if used with the latter goal, results can be deceiving due to potential strong projection effects along the z-axis.


 As can be clearly seen from the colour gradient and from the partial correlation coefficient arrow in Fig.\ref{fig:1}, the most important quantity responsible for increasing the star formation rate surface density is the molecular gas mass surface density - this reveals that the Schmidt-Kennicutt relation is the primary relation driving $\Sigma_{\rm SFR}$ for star forming regions, as previously found in the ALMaQUEST data by \citet{2020EllisonSFEMNRAS.493L..39E}. 
 Once the dependence on $\Sigma_{\rm H_2}$ is taken into account, Figure \ref{fig:1} shows that $\Sigma_*$ plays little or no role in determining the $\Sigma_{\rm SFR}$.

The contours in Figure \ref{fig:1} also illustrate the relation between  $\Sigma_*$ and  $\Sigma_{\rm H_2}$, which has already been identified as the resolved Molecular Gas Main Sequence by \citet{Lin2019}. We have fitted it with an orthogonal distance regression (ODR) in log-log, shown by the blue line, with slope $\sim 1$, specifically:

\begin{equation}
\log{(\Sigma _{\rm H_2})} = 1.04~(\pm0.01)~\log{(\Sigma_*)}-1.44~ (\pm0.06)~.
\end{equation}

We also mention, for completeness, that with our ALMaQUEST data the Schmidt-Kennicutt relation has the form:

\begin{equation}
\log{(\Sigma _{\rm SFR})} = 1.19~(\pm 0.01)~\log{(\Sigma_{\rm H_2})}-10.41~(\pm 0.05)~.
\end{equation}

The combination of these relations in figure \ref{fig:1} reveals that the resolved Star Formation Main Sequence, i.e. the relation between  $\Sigma_{\rm SFR}$ and $\Sigma_*$, is not an intrinsic property of galaxies, rather it stems from the Schmidt-Kennicutt relation (dependence of $\Sigma_{\rm SFR}$ on $\Sigma_{\rm H_2}$) and from the rMGMS (dependence of $\Sigma_{\rm H_2}$ on $\Sigma_*$), i.e. in confirmation of \citet{Lin2019}.


\begin{figure}
    \centering
    \includegraphics[width=\columnwidth]{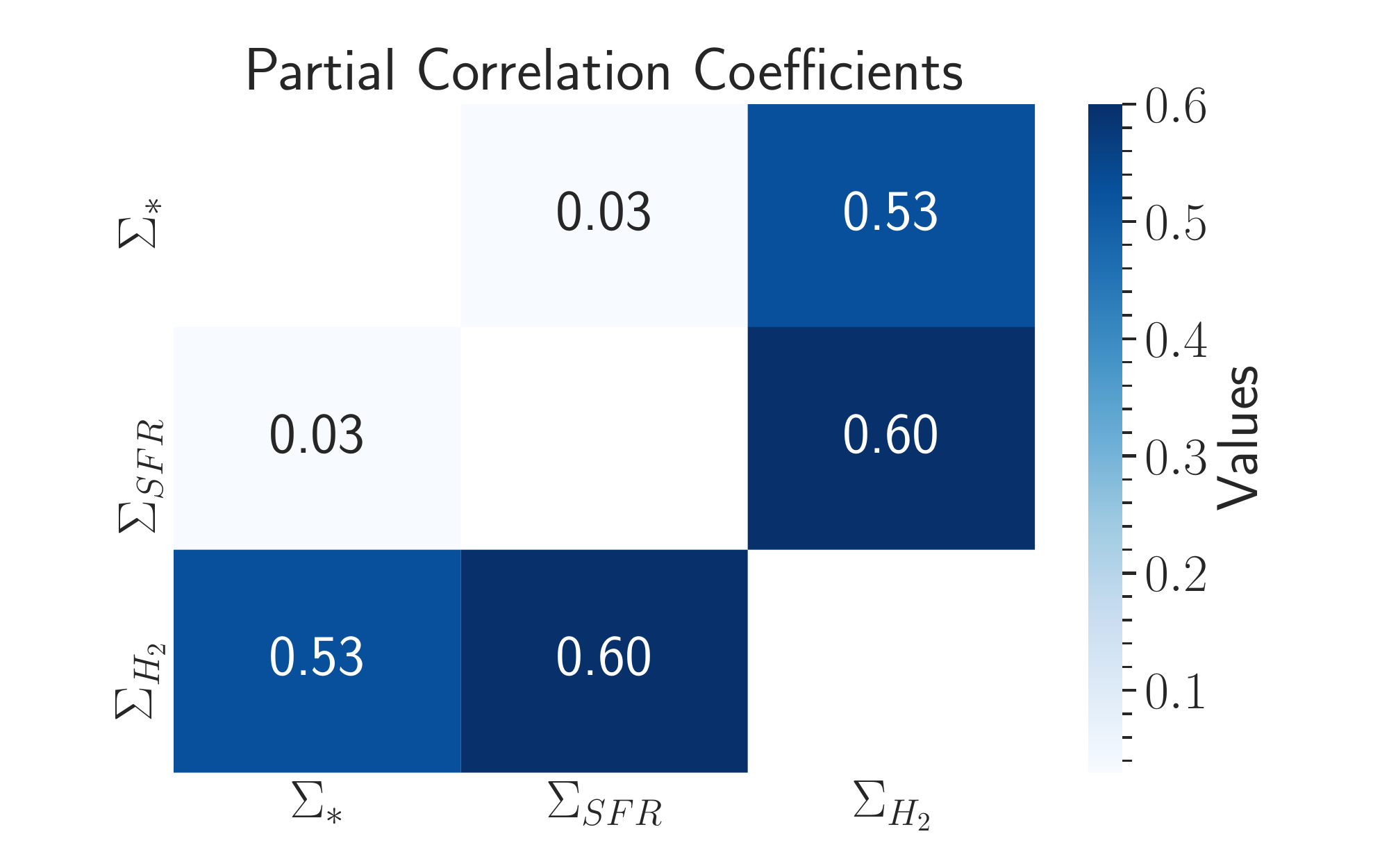}
    \caption{Partial Correlation Coefficients between $\Sigma_*$, $\Sigma_{\rm SFR}$ and $\Sigma_{\rm H_2}$ with the third quantity held constant. The colour coding shows the strength of the partial correlation coefficients. Errors obtained by bootstrap random sampling are of the order $\sim 0.01$ or less.}
    \label{fig:PCCs}
\end{figure}



Figure \ref{fig:PCCs} shows these results simultaneously and more quantitatively by giving the Partial Correlation Coefficient values for the three quantities in pairs, whilst controlling the third quantity. 
The partial correlation coefficient between $\Sigma_{\rm H_2}$ and $\Sigma_{\rm SFR}$ is the largest, followed by that between $\Sigma_*$ and $\Sigma_{\rm H_2}$.  From this it can be inferred that the resolved Schmidt-Kennicutt relation is the strongest, closely followed by the rMGMS. The partial correlation coefficient between $\Sigma_*$ and $\Sigma_{\rm SFR}$ is nearly zero, which clearly demonstrates that the $\Sigma_{\rm SFR}$ - $\Sigma_*$ relation is not fundamental, but rather arises out of the other two more fundamental scaling relations. Errors for the partial correlation coefficients, obtained by bootstrap random sampling, are of the order $\sim0.01$ or less.

These results unambiguously confirm the result of \cite{Lin2019} and \citet{2020EllisonSFEMNRAS.493L..39E}, that the Star Forming Main Sequence is a by-product of the Schmidt-Kennicutt relation combined with the Molecular Gas Main Sequence on resolved scales, but  more robustly and with high statistical significance.



\subsection{Random Forest Regression Results}
We also run a Random Forest regression analysis in order to identify the most important parameter for determining the star formation rate surface density. 
As correlation does not necessarily imply causation, random forest regression provides an important check on correlation results, especially as the random forest can find other links that may be missed.

Random Forest regression is a form of machine learning where parameter importances can be predicted by using many different decision trees. The decision trees work via trying to reduce the Gini Impurity \citep{2012PedragosaarXiv1201.0490P} at each stage of the process. A target quantity is selected and removed from the data, leaving the features that could contribute to it. These two datasets, the target and the features, are then split into a training sample and a validation sample. The algorithm is applied to the training sample which sorts the data into different nodes in several trees in order to minimise Gini Impurity. This creates a model which can then be applied to the validation sample in order to determine the parameter importances \citep[e.g.][]{2020Bluck,2020BluckB}. 
Random forest regression is particularly useful as it does not require monotonicity and can uncover highly non-linear trends. Moreover, it can simultaneously explore the dependence on multiple inter-correlated quantities \citep[for more details on the Random Forest technique see][]{Bluck2021submitted}.

We use the random forest to investigate what is the most predictive quantity of $\Sigma _{\rm SFR}$ among the following:
the molecular gas mass surface density ($\Sigma_{\rm H_2}$), the stellar mass surface density ($\Sigma_*$), the total molecular gas mass ($\rm M_{H_2}$), the total stellar mass ($\rm M_*$), the metallicity ($Z$), the galactocentric distance ($D$), and a random variable (R) (selected from a uniform  distribution).

We check the performance of the random forest for both the training and test samples to ensure we avoid overfitting the training data. We do this by plotting the $\Sigma_{\rm SFR}$ target against predicted for both the training and testing samples and comparing their mean squared errors (MSE). By fine-tuning the hyper-parameters using a randomized cross search validation method we minimise the difference between the MSE of the train and test samples.

\begin{figure}
    \centering
    \includegraphics[width=\columnwidth]{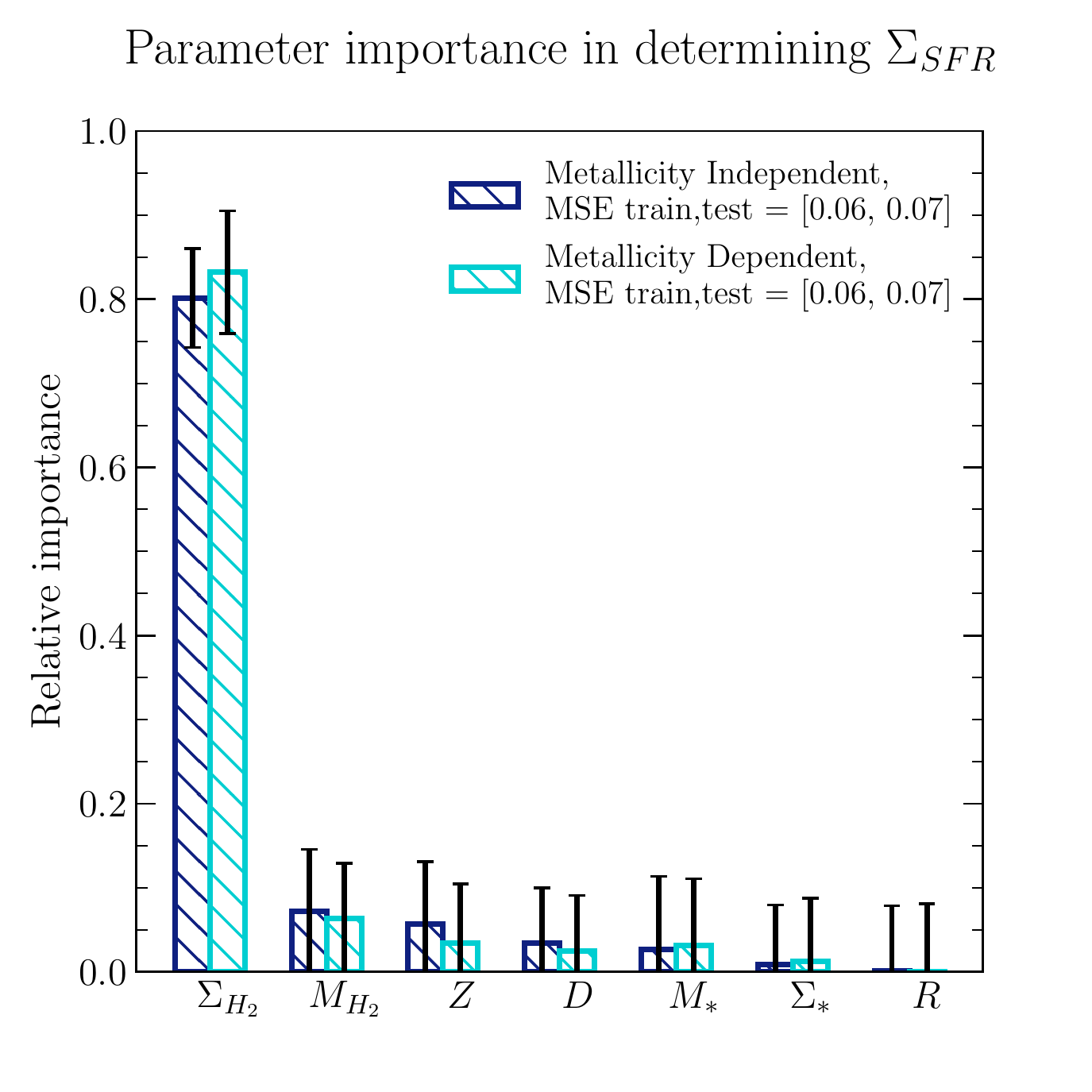}
\caption{Bar-chart showing the relative importances of each parameter in determining the SFR surface density ($\Sigma _{\rm SFR}$), from a Random Forest analysis, for the case of both metallicity-independent and metallicity-dependent molecular gas conversion factor. The error-bars are obtained by bootstrap random sampling. The mean squared errors (MSE) are shown for both the testing and the training (validation) set. Clearly, $\Sigma_{H_2}$ is the most important parameter for predicting $\Sigma_{\rm SFR}$.}
\label{fig:Import}
\end{figure}

Figure \ref{fig:Import} shows the parameter's importance in determining $\Sigma_{\rm SFR}$ for both the metallicity independent (dark blue) and metallicity dependent (light blue) CO conversion factor. Figure \ref{fig:Import} demonstrates that the random forest regression unambiguously identifies the molecular gas mass surface density, $\Sigma_{\rm H_2}$, as the most important parameter for determining the star formation rate surface density, in agreement with the partial correlation coefficients results (see \citet{2020EllisonSFEMNRAS.493L..39E} for similar results using an artificial neural network). All other quantities have far lower importance. In particular, $\Sigma_{*}$ is consistent with the importance of a totally random variable, i.e. $\Sigma_{*}$ is consistent with being intrinsically totally unimportant in determining the $\Sigma_{\rm SFR}$, once its dependence on $\Sigma_{\rm H_2}$ is taken into account.
This clearly demonstrates that \emph{the resolved star forming main sequence is not a fundamental scaling relation.}

\subsection{Differential Measurement Uncertainty Test}


A possible cause of error with this analysis would be that the Partial Correlation Coefficients and Random Forest regression could have been biased by certain quantities having more or less measurement uncertainty than others. An example of this being if $\Sigma_{*}$ had significantly more uncertainty than $\Sigma_{\rm H_2}$ and $\Sigma_{\rm SFR}$, then a weaker correlation would be found between $\Sigma_{*}$ and $\Sigma_{\rm SFR}$ than that between $\Sigma_{\rm H_2}$ and $\Sigma_{\rm SFR}$, without this being a true reflection of the intrinsic correlations. In order to investigate this possibility we performed a differential measurement uncertainty test, which involves observing how much the noise of one of the measurements needs to be artificially boosted, in order to invert the previously observed trend.

\begin{figure}
    \centering
    \includegraphics[width=\columnwidth]{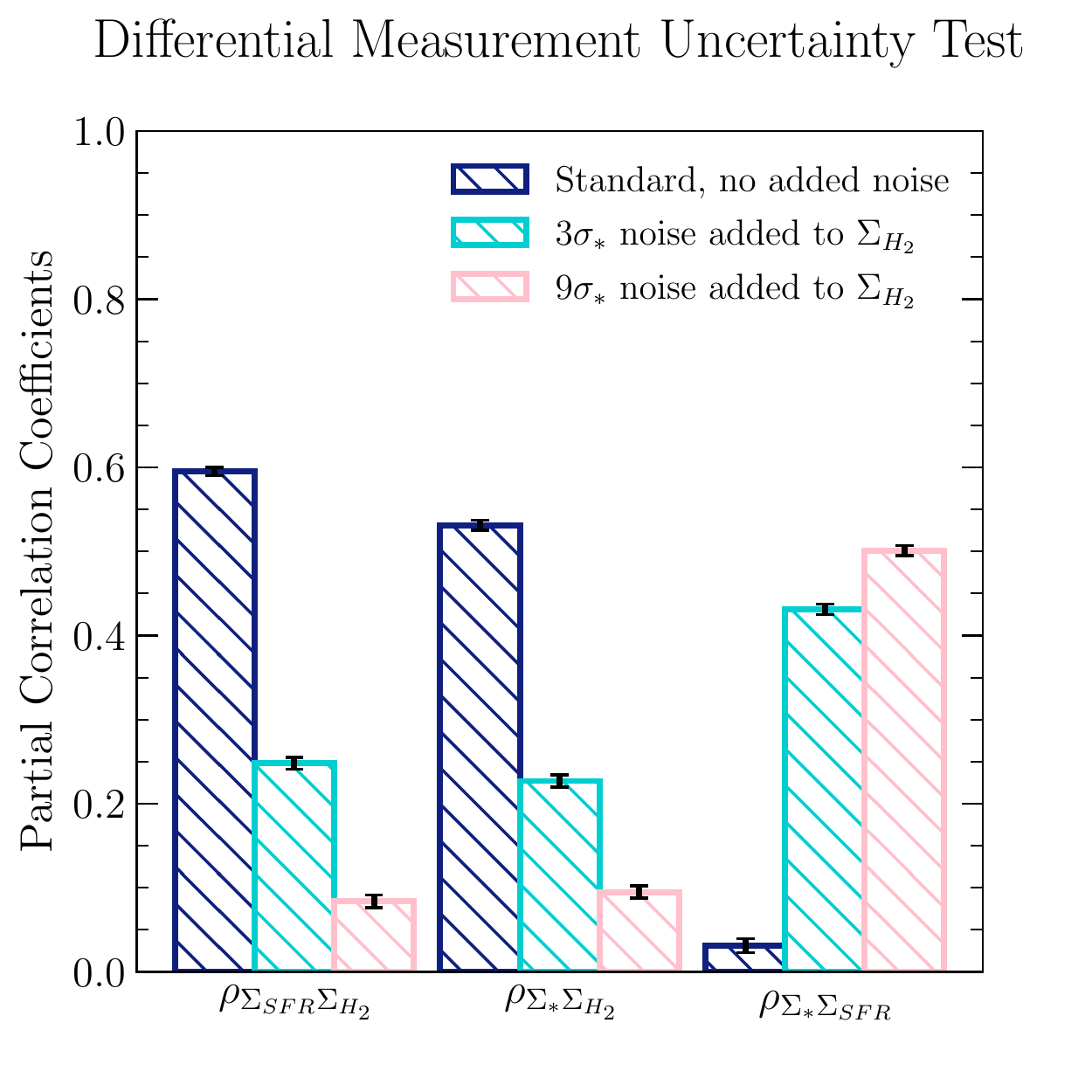}
    \caption{The partial correlation coefficients between the three quantities, $\Sigma_*$, $\Sigma_{\rm SFR}$, and $\Sigma_{\rm H_2}$ , where $\rho_{XY}$ is the partial correlation coefficient between quantities $X$ and $Y$, controlling for $Z$. The three scenarios outlined are: the standard case, as in figure \ref{fig:PCCs}, and after the addition of 3$\sigma_*$, and $9\sigma_*$, worth of Gaussian random noise to $\Sigma_{H_2}$.  }
    \label{fig:PCCs_noise}
\end{figure}

We artificially add Gaussian random noise to (just) $\Sigma_{\rm H_2}$, by increasing $\sigma$ until the partial correlation between $\Sigma_{*}$ and $\Sigma_{\rm SFR}$ reaches the previous level of that between $\Sigma_{\rm H_2}$ and $\Sigma_{\rm SFR}$. 
 We add Gaussian noise in units of $\sigma_*$, i.e. the uncertainty affecting $\Sigma_*$, which we conservatively assume to be 0.2 dex \citep[][]{Mendel2014ApJS..210....3M}.
We find that even adding $9\sigma_*$ additional noise to $\Sigma_{H_2}$ the partial correlation coefficient between $\Sigma_{*}$ and $\Sigma_{\rm SFR}$ does not reach that of the previous value between $\Sigma_{\rm H_2}$ and $\Sigma_{\rm SFR}$.
This is illustrated in Figure \ref{fig:PCCs_noise}, which shows the differential measurement uncertainty test results. Plotted are the partial correlation coefficient results between $\Sigma_*$, $\Sigma_{\rm SFR}$, and $\Sigma_{\rm H_2}$ (where $\rho _{XY}$ is the partial correlation coefficient between quantities $X$ and $Y$, controlling for $Z$), in the case of: no added noise (giving obviously the same result as figure \ref{fig:PCCs}); 3$\sigma_*$ of Gaussian random noise added to $\Sigma_{\rm H_2}$; and  9$\sigma_*$ of Gaussian random noise added to $\Sigma_{\rm H_2}$. 
The errors are obtained by bootstrap random sampling (100 times with the sample size equal to the dataset). As can be seen in figure \ref{fig:PCCs_noise}, it requires at least 9$\sigma_*$ of Gaussian random noise to be added to $\Sigma_{\rm H_2}$ in order to completely invert the trend. Even so, the Partial Correlation Coefficient between $\Sigma_{\rm SFR}$ and $\Sigma_*$ (controlling for $\Sigma_{H_2}$), in the case of $\Sigma_{H_2}$ having errors so enormously boosted, (rightmost pink bar) has not yet reached the same value obtained for the Partial Correlation Coefficient between $\Sigma_{\rm SFR}$ and $\Sigma_{H_2}$ (controlling for $\Sigma_*$) without boosting of the errors (leftmost dark blue bar).

This can be interpreted as follows: in order to disrupt the strength of the correlation between $\Sigma_{\rm SFR}$ and $\Sigma_{\rm H_2}$, and make it as weak as the (lack of) correlation between $\Sigma_{\rm SFR}$ and $\Sigma_{*}$, we need to artificially increase the uncertainty of $\Sigma_{\rm H_2}$ by the order of 9 times the uncertainty of $\Sigma_*$. Since it is implausible that the uncertainty on $\Sigma_{\rm H_2}$ has been underestimated by such a large factor, this result strongly disfavours the possibility of uncertainty contributing to skewing the partial correlation results.

We also performed the differential measurement test for the Random Forest regression and again found that, in order for $\Sigma_*$ to obtain the same importance as $\Sigma_{\rm H_2}$ in predicting $\Sigma_{\rm SFR}$, we needed to add a minimum of $9\sigma_*$ of extra uncertainty to $\Sigma_{\rm H_2}$. This provides another independent test of the partial correlation coefficient results.


\section{Discussion}
Our primary goal is determining the relative importance and origin of the resolved scaling relations for star forming regions. We found that the strongest scaling relation is the resolved Schmidt-Kennicutt relation between $\Sigma_{\rm H_2}$ and $\Sigma_{\rm SFR}$, which follows from molecular gas fuelling star formation: the greater the amount of molecular gas available in a region of a galaxy, the greater the star formation in that region.  
We found that the partial correlation coefficient for this relationship is $\rho=0.60$ (between $\Sigma_{\rm H_2}$ and $\Sigma_{\rm SFR}$ whilst controlling for $\Sigma_*$), making it the largest.
We also found that the Random Forest regression clearly identifies $\Sigma_{\rm H_2}$ as the most important parameter, by far, in contributing to the determination of $\Sigma_{\rm SFR}$, instead of $\Sigma_*$ and several other variables of interest.. The Random Forest regression is useful for adding further confirmation. As has been stated many times before, correlation does not necessarily imply causation.
Including Random Forest regression into our analysis, alongside Partial Correlation Coefficients, enables us to get closer to testing the causality of this relationship \citep[see][for various tests on the extraction of causation from correlation using the RF technique.]{Bluck2021submitted} 

The resolved Molecular Gas Main Sequence was found to be the second strongest relationship, via the Partial Correlation Coefficients, with a value of $\rho=0.53$ between $\Sigma_{\rm H_2}$ and $\Sigma_*$. Explaining the causality in this case is more complicated. It is apparent that, if there is causality here, it is $\Sigma_{\rm H_2}$ that depends upon $\Sigma_*$, rather than the inverse. A possible explanation for this could be, as was suggested in \citet{Lin2019}, that the molecular gas is following the gravitational potentials of the galaxy. The stellar mass surface density ($\Sigma_*$) would either be the cause of these gravitational potentials (as in the case of small dark matter content), or would itself follow the gravitational potentials (possibly dominated by dark matter) alongside $\Sigma_{\rm H_2}$ inducing the correlation. It would therefore be interesting to explore whether the dependence of $\Sigma_{\rm H_2}$ is really upon the stellar mass (as in the first scenario), the dynamical mass (as in the second) or a separate quantity.

The resolved Star Forming Main Sequence is found to simply be a by-product of the resolved Schmidt-Kennicutt relation and the resolved Molecular Gas Main Sequence. This is supported by the nearly null partial correlation coefficient between $\Sigma_*$ and $\Sigma_{\rm SFR}$, whilst controlling for $\Sigma_{\rm H_2}$, of $\rho=0.03$. Further evidence comes from the Random Forest regression results, where $\Sigma_*$ is found to have no role in determining $\Sigma_{\rm SFR}$ (see figure \ref{fig:Import}). Therefore, we find that the star formation rate does not causally depend on the stellar mass surface density. This provides additional statistical evidence to support the conclusions of \citet{Lin2019} and \cite{2020EllisonSFEMNRAS.493L..39E}.

The differential measurement uncertainty test enables us to rule out uncertainties in the measurements as significant contributing factors to our results. This ensures that the Partial Correlation Coefficient results and Random Forest regression have not been skewed by noisy data. Our analysis shows that this scenario is heavily disfavoured for our results.

One area of uncertainty concerns the overlap of quality measurements. Spaxels with strong CO flux can have poor H$\alpha$ flux (hence poor SFRs), whilst spaxels with strong H$\alpha$ flux  can have poor CO flux.
An area for further analysis would be what effect this overlap of quality measurements has on our results. This could be investigated by stacking spectra for the undetected regions.
We do note, however, that this should not be a major issue with our analysis, as we are primarily selecting spaxels on the Main Sequence. 

An interesting point of discussion is whether these results also apply to the integrated, global properties (M$_*$, SFR, and M$_{H_2}$). This aspect is being explored by combining data from multiple surveys and the results will be presented in a forthcoming paper.

We finally mention that \citet{Dou2020}  claim that the more fundamental scaling relations are those that exist between the gas fraction ($\rm M_{gas}/M_*$) and specific star formation rate ($\rm sSFR=SFR/M_*$), which, as a consequence leads to relations of them both with star formation efficiency ($\rm SFE=SFR/M_{gas}$). These relations could then have counterparts on resolved scales. Since these relations do not involve directly observed quantities, rather a combination of $\Sigma_{\rm H_2}$, $\Sigma_{\rm SFR}$, and $\Sigma_{*}$, it is more difficult to test these relations with our methodology. However, these scaling relations will be explored in a separate paper.

\section{Conclusions}


We use data from the MaNGA and ALMaQUEST surveys to explore the spatially resolved scaling relations between $\Sigma_{\rm SFR}$, $\Sigma_{\rm H_2}$ and $\Sigma_{*}$. We improve upon previous work by employing Partial Correlation Coefficients and a Random Forest regression analysis. Both of these techniques are highly effective at extracting causality from inter-correlated data. 

We summarise our results as follows:
\begin{enumerate}
  \item The Partial Correlation Coefficient analysis shows that $\Sigma_{\rm SFR}$ depends directly on just $\Sigma_{\rm H_2}$ (resolved Schmidt-Kennicutt relation) and has no intrinsic dependence on $\Sigma_*$ (resolved Star Forming Main Sequence).
  
  \item We confirm the existence of a resolved Molecular Gas Main Sequence (rMGMS) between $\Sigma_{\rm H_2}$ and $\Sigma_{*}$.
  
  \item Our Random Forest regression analysis confirms that the most important parameter driving $\Sigma_{\rm SFR}$ is $\Sigma_{\rm H_2}$, while all other galactic parameters have far less importance. In particular, $\Sigma_{*}$ does not play a role. 
  
  \item Therefore, the resolved Star Forming Main Sequence is not an intrinsic property of galaxies, it is simply an emergent by-product of the resolved Schmidt-Kennicutt relation and the resolved Molecular Gas Main Sequence.
  
  \item We rule out differential measurement uncertainty as a plausible origin of these results.
  
\end{enumerate}

\section*{Acknowledgements}
W.B., R.M. and A.B. acknowledge support by the Science and Technology Facilities Council (STFC) and ERC Advanced Grant 695671 "QUENCH".

The authors would like to thank the staffs of the East-Asia and North-America ALMA ARCs for their support and continuous efforts in helping produce high-quality data products. This paper makes use of the following ALMA data:\\
ADS/JAO.ALMA\#2015.1.01225.S,
ADS/JAO.ALMA\#2017.1.01093.S, 
ADS/JAO.ALMA\#2018.1.00541.S,
\\and ADS/JAO.ALMA\#2018.1.00558.S. 

ALMA is a partnership of ESO (representing its member states), NSF (USA) and NINS (Japan), together with NRC (Canada), MOST and ASIAA (Taiwan), and KASI (Republic of Korea), in cooperation with the Republic of Chile. The Joint ALMA
Observatory is operated by ESO, AUI/NRAO and NAOJ.

Funding for the Sloan Digital Sky Survey IV has been provided by the Alfred P. Sloan Foundation, the U.S. Department of Energy Office of Science, and the Participating Institutions. SDSS acknowledges support and resources from the Center for High-Performance Computing at the University of Utah. The SDSS web site is www.sdss.org.

SDSS is managed by the Astrophysical Research Consortium for the Participating Institutions of the SDSS Collaboration including the Brazilian Participation Group, the Carnegie Institution for Science, Carnegie Mellon University, Center for Astrophysics | Harvard \& Smithsonian (CfA), the Chilean Participation Group, the French Participation Group, Instituto de Astrofísica de Canarias, The Johns Hopkins University, Kavli Institute for the Physics and Mathematics of the Universe (IPMU) / University of Tokyo, the Korean Participation Group, Lawrence Berkeley National Laboratory, Leibniz Institut für Astrophysik Potsdam (AIP), Max-Planck-Institut für Astronomie (MPIA Heidelberg), Max-Planck-Institut für Astrophysik (MPA Garching), Max-Planck-Institut für Extraterrestrische Physik (MPE), National Astronomical Observatories of China, New Mexico State University, New York University, University of Notre Dame, Observatório Nacional / MCTI, The Ohio State University, Pennsylvania State University, Shanghai Astronomical Observatory, United Kingdom Participation Group, Universidad Nacional Autónoma de México, University of Arizona, University of Colorado Boulder, University of Oxford, University of Portsmouth, University of Utah, University of Virginia, University of Washington, University of Wisconsin, Vanderbilt University, and Yale University.


\section*{Data Availability}
 
The ALMA data used is publicly available through the ALMA archive http://almascience.nrao.edu/aq/.

The MaNGA data that is used in this work is publicly available at 
https://www.sdss.org/dr15/manga/manga-data/.

\bibliographystyle{mnras}
\bibliography{example} 




\appendix

\bsp	
\label{lastpage}
\end{document}